# Bridging Quantum Computing Paradigms toward Semiconductor Yield: A Controlled CV-versus-DV Comparison on Wafer-Map Defect Classification

Yeonhong Kim[1,2], Jonghyeok Im[2], Monu Nath Baitha[3,*] and Kyoungsik Kim[2,*]

[1]Department of Materials Science and Engineering, Yonsei University, Seoul 03722, Republic of Korea.

[2]Department of Mechanical Engineering, Yonsei University, Seoul 03722, Republic of Korea.

[3]School of Electronics Engineering, Vellore Institute of Technology, Vellore 632014, India.

[*]Corresponding E-mails: kks@yonsei.ac.kr, monu.baitha@vit.ac.in.

## Abstract

Realizing quantum neural networks (QNNs) in industry requires knowing which quantum computing paradigm suits which task. Motivated by AI accelerators and high-bandwidth memory, where die stacking makes wafer-level defect screening central to yield, we study WM-811K wafer-map defect classification (eight classes), comparing the dominant paradigms, continuous-variable (CV) and discrete-variable (DV), under controlled conditions. To isolate the quantum circuit as the sole variable, a shared convolutional backbone (~4.3M parameters) feeds interchangeable heads (classical dense, CV-QNN, or DV-QNN) as the only structural difference; each quantum head is scaled over three sizes (3, 4, 8 qumodes/qubits). The CV head consistently outperforms the DV head: at four qumodes/qubits it reaches 79.7 ± 1.8% accuracy versus 61.6 ± 1.4%, a non-overlapping 18-point gap. The advantage is sharpest on the spatially localized Edge-Loc class, easily confused with Scratch, which CV recovers with recall 0.66 ± 0.06 while DV fails at every size (≤0.05), showing the structured CV layer better captures fine spatial distinctions between defect types. Training curves show the DV limitation is a representational-capacity ceiling, not an optimization failure; at the Fock cutoff used here (d = 2) the CV advantage reflects two intrinsic properties, a structured, neural-network-analogue layer and continuous phase-space encoding, not Hilbert-space dimensionality. On IBM hardware, DV accuracy holds at shallow depth, degrading only at the deepest circuit. Both quantum heads remain below the classical baseline (85.0%), but the controlled setting isolates where a structured head already helps and, as noise and scale improve, which paradigm can deliver practical advantage.

## 1 Introduction

As semiconductor process nodes shrink toward the single-nanometer regime, patterned-feature density rises and process margins tighten, so that subtle process excursions produce increasingly intricate spatial defect signatures on the wafer map (Wu et al. 2015; Saqlain et al. 2020). This pressure is amplified by the rapid growth of artificial-intelligence accelerators: based on 2025 AI-chip shipments, TrendForce estimates that high-bandwidth memory (HBM) demand will rise by more than 130% year-over-year in 2025 and continue to grow by over 70% in 2026 (TrendForce 2025). Because such memory stacks many dies vertically, a single defective die can compromise an entire package, so wafer-level defect screening remains central to yield and cost. Accurate, automated classification of wafer-map defect patterns, which encodes the spatial signature of process faults, is therefore an essential tool for root-cause analysis and yield improvement in modern manufacturing (Jeong et al. 2023), motivating continued interest in stronger pattern-recognition methods.

Quantum machine learning (QML) offers an alternative computational substrate that may benefit such tasks (Biamonte et al. 2017), with growing interest in image-classification applications (Senokosov et al. 2024). Quantum neural networks (QNNs) come in two paradigms: discrete-variable (DV) circuits on qubits with $2^N$-dimensional Hilbert spaces, and continuous-variable (CV) circuits on qumodes, whose Fock space is formally infinite-dimensional but is truncated to a finite cutoff for classical simulation (here $d = 2$ per qumode; see Section 5.2), processed through Gaussian and non-Gaussian gates (Killoran et al. 2019a). These paradigms differ in their computational primitives, with potential consequences for classification.

Prior work leaves the comparison unsettled. Yang and Sun (2022) applied a DV circuit atop an SP&A-Net backbone to WM-811K (98.10%), but the heavy backbone dominates performance and masks the quantum contribution. Lopez et al. (2025) compared CV and DV on MedMNIST with PCA features and small circuits, reporting a CV advantage; Choe (2022) demonstrated CV-QNN on MNIST. To our knowledge, no prior study isolates the quantum head on an industrial dataset while preserving spatial structure.

We address the following questions: under a strictly controlled protocol (shared backbone, interchangeable heads, matched parameter budget), how do CV and DV quantum heads compare on WM-811K, and what design principles follow? Our contributions are:

- **A controlled CV-versus-DV comparison on an industrial dataset.** All models share an identical convolutional backbone (~4.28M parameters) that performs feature extraction up to a common 512-dimensional representation; only the downstream classification head differs, classical dense, CV-QNN, or DV-QNN. Because feature-extraction capacity is held fixed by this shared backbone, any difference in performance is attributable to the quantum head rather than to the backbone. This makes the quantum circuit the sole structural variable and avoids the "masking" of the quantum contribution by a dominant classical backbone that limits prior wafer-map studies (Yang and Sun 2022).
- **The CV head consistently outperforms the DV head, decisively on spatial classes.** Over three runs at four qumodes/qubits, CV reaches $79.7 \pm 1.8\%$ versus DV's $61.6 \pm 1.4\%$, and on the spatially localized Edge-Loc class CV recovers recall $0.66 \pm 0.06$ while DV essentially fails ($\leq 0.05$ at every size, with the four-qubit head at $0.049 \pm 0.020$). We trace this to two intrinsic properties of the CV head, its structured (NN-analogue) layer design and its continuous phase-space encoding, versus the unstructured, discretely encoded DV ansatz.
- **A theoretical clarification.** At the cutoff used here ($d = 2$), the CV model is not genuinely non-Gaussian; the CV advantage is best understood as arising from continuous phase-space encoding and structured design, not from Hilbert-space dimensionality. This corrects a common assumption in the CV-QNN literature.

- **Practical hardware and interface findings.** DV-QNN classification accuracy is preserved at shallow depth on IBM hardware (8–12 two-qubit gates, $\Delta \approx +0.01$) and degrades at the deepest circuit (28 two-qubit gates, $\Delta = -0.07$), and DV training requires a deterministic classical–quantum encoding interface.

## 2 Background and Related Work

### 2.1 Continuous-Variable Quantum Neural Networks

The CV-QNN framework of Killoran et al. (2019a) defines a network layer using Gaussian and non-Gaussian gates that mirror a classical layer $L(x) = \Phi(Wx + b)$. The linear map W is realized by interferometers and squeezing ($D \circ U_2 \circ S \circ U_1$); by SVD, $W = O_2 \Sigma O_1$. The bias b is a displacement $D(\alpha)$, and the nonlinearity $\Phi$ is a Kerr gate $K(\kappa)|n\rangle = e^{ikn^2}|n\rangle$. This explicit "affine-then-nonlinearity" correspondence is the architectural feature we will argue underlies the CV head's stable trainability. We note a caveat developed in Section 5.2: the universality of W requires a full interferometer mesh ($m(m-1)/2$ beamsplitters), and the non-Gaussian character of the Kerr gate is only realized for Fock cutoff $d \geq 3$.

### 2.2 Discrete-Variable Quantum Neural Networks

DV-QNN operates on qubits using parameterized rotations $R_Y(\theta)$, $R_Z(\varphi)$ and CNOT entanglers, with data re-uploading (Pérez-Salinas et al. 2020) interleaving encoding and variational layers for universal approximation. Unlike the CV layer, the DV circuit is a generic hardware-efficient ansatz: it carries no explicit correspondence to a neural-network layer, and its trainability is known to depend on initialization, depth, and entangling structure, with barren-plateau risks for poorly conditioned ansätze (McClean et al. 2018; Sim et al. 2019). This structural difference, rather than any single gate, motivates our comparison.

### 2.3 WM-811K Wafer Defect Classification

WM-811K (Wu et al. 2015) contains >811,000 production wafer maps across eight defect classes. Classical state-of-the-art exceeds 95% with advanced backbones (Saqlain et al. 2020); our aim is not to match it but to compare quantum heads under a controlled, deliberately simple backbone.

Each of the eight classes corresponds to a distinct spatial signature and an associated process root cause (Table 1). Correct classification is a prerequisite for accurate root-cause diagnosis, which in turn supports timely corrective action and yield-loss mitigation. This consequence is especially acute for the Edge-Loc and Scratch classes, whose morphologies share a superficial similarity (elongated clusters near or along the wafer edge) yet whose root causes call for entirely different interventions, thermal-process tuning versus mechanical-handling inspection (Shin et al. 2024; Hansen et al. 1997). Misclassifying one as the other therefore misdirects the fab's corrective response toward the wrong process module, compounding yield loss rather than mitigating it; we return to this point in Sections 4.2 and 5.5.

**Table 1.** Defect patterns in the WM-811K dataset and their associated process root causes. Adapted from Shin et al. (2024); primary root-cause references as cited therein.

| Defect class | Pattern description | Process root cause |
|---|---|---|
| Center | Defects concentrated in the wafer center | Irregular RF operation or unusual liquid flow |
| Donut | Ring-shaped pattern with an opening in the middle | Residue resistant to removal during photoresist cleaning |
| Edge-Loc | Defects clustered on a localized arc of the wafer edge | Irregular temperature annealing or load-lock valve impurities (Hansen et al. 1997) |
| Edge-Ring | Defects along the entire wafer periphery | Anomalous temperature regulation during RTP or cold process |
| Loc | Defects clustered in an interior region | Vacuum-pressure differences from slit-valve leaks, poor pump operation, or vibration |
| Near-full | Defects covering nearly the entire wafer | Photoresist rupture from electron overcharge in plasma ion implant |
| Scratch | Continuously connected linear defect trail | Mechanical damage by transfer robots, handling, or CMP |
| Random | Sporadically distributed defects | Abnormal vacuum, gas, or fluid eruption |

# 3 Methods

## 3.1 Dataset and Preprocessing

WM-811K is preprocessed into eight classes at 64×64 resolution, stratified-split before augmentation. Training: 4,800 samples (600/class, balanced via horizontal/vertical flips and ±2 px translation). Validation/test: 2,552 samples each (original imbalanced distribution). Rotation augmentation was excluded after it was found to destroy the Edge-Loc/Edge-Ring spatial distinction (Yu et al. 2023). Normalization uses training-set statistics.

## 3.2 Controlled Evaluation Protocol

All models share a strictly identical CNN backbone (Fig. 1a): three convolutional blocks (32→64→128 filters, 3×3, ReLU, MaxPool, Dropout) followed by Dense(8192→512, ELU), terminating at a 512-dimensional feature vector. Each head then transforms this vector:

- **Classical head:** Dense(512→16, ELU) → Dense(16→8, softmax)
- **CV-QNN head:** Dense(512→d, ELU) → clip(±1.0) → CV circuit → Dense(out→8, softmax)
- **DV-QNN head:** Dense(512→d, ELU) → per-sample feature normalization → bounded angle encoding (data re-uploading) → DV circuit → Dense(out→8, softmax)

The DV encoding interface (Section 3.4) is deterministic per sample: it applies per-sample normalization and a bounded angle map, and, critically, contains no stochastic regularization (dropout) immediately before the quantum circuit. We found this necessary for DV trainability (Section 5.4). Total parameters are held near 4.3M (Table 2).

**Table 2.** Model configurations. In the CV-QNN rows, *m* denotes the number of qumodes and *c* the Fock cutoff dimension per qumode (e.g., "3m, c=3" is three qumodes at cutoff 3); in the DV-QNN rows, the size is given in qubits. "Encoding dimension" is the dimension of the vector passed from the classical interface into the quantum circuit; "Circuit params" counts the trainable quantum-circuit (variational) parameters; "Measurement dimension" is the size of the circuit's measured output vector. The shared backbone (~4.28M) is identical across models; differences arise only in the head. The DV head additionally uses ~90 encoding-conditioning parameters (batch normalization + learnable scale), negligible against the total.

| Model | Interface params (512→encoding) | Encoding dimension | Circuit params | Measurement dimension | Total model params |
|---|---|---|---|---|---|
| Classical CNN | 8,208 | 16 | 0 | — | ~4.30M |
| CV-QNN (3m, c=3) | 11,286 | 22 | 138 | 27 | ~4.30M |
| CV-QNN (4m, c=2) | 15,390 | 30 | 190 | 16 | ~4.30M |
| CV-QNN (8m, c=2) | 31,806 | 62 | 398 | 256 | ~4.32M |
| DV-QNN (3q) | 15,390 | 30 | 24 | 8 | ~4.30M |
| DV-QNN (4q) | 15,390 | 30 | 32 | 16 | ~4.30M |
| DV-QNN (8q) | 15,390 | 30 | 64 | 256 | ~4.30M |

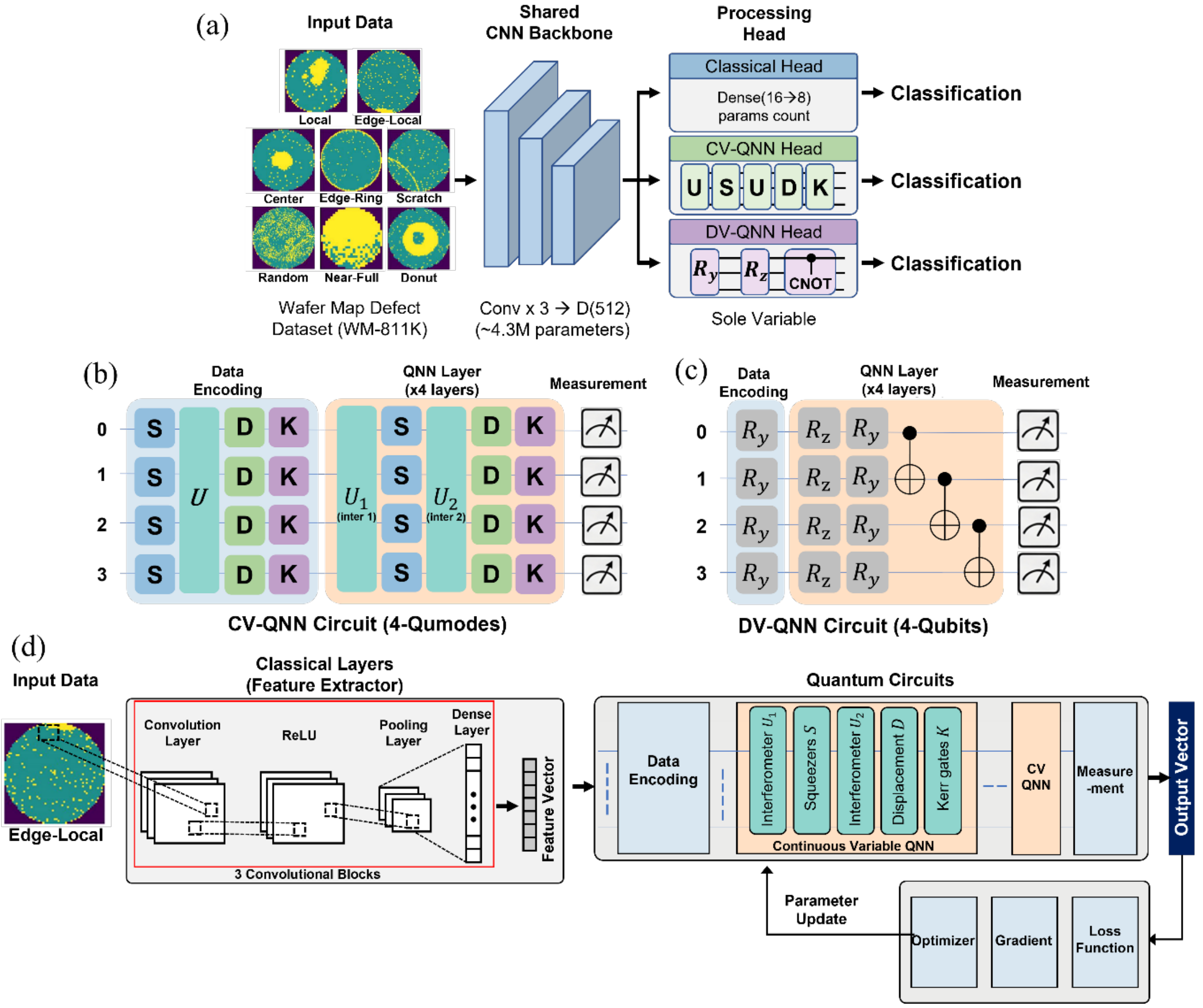

**Fig. 1 Controlled protocol and architectures** - (a) shared-backbone protocol with three interchangeable heads; (b) CV-QNN circuit (4-qumode S→BS→R→D→K + 4 layers); (c) DV-QNN circuit (deterministic per-sample encoding → $R_Y$ → $R_Z$ + $R_Y$ → CNOT chain × 4); (d) variational parameter-update loop for the CV-QNN — circuit measurements feed a classical optimizer that updates the gate parameters (displacement, squeezing, rotation, Kerr), with the shared CNN backbone supplying the encoded

input vector.

### 3.3 CV-QNN Circuit Design

Following Killoran et al. (2019a) and Choe (2022), the CV-QNN (Fig. 1b) comprises an encoding circuit (Squeezing → Beamsplitter interferometer → Rotation → Displacement → Kerr) and four variational layers ($U_1$ → S → $U_2$ → D → K). Output is the Fock-basis probability distribution of dimension $(\text{cutoff})^m$. A clip-by-value (±1.0) layer conditions the classical features entering the gates, keeping the Fock population concentrated in the low-photon subspace at cutoff 2. The implications of the d = 2 truncation for non-Gaussianity are analyzed in Section 5.2.

### 3.4 DV-QNN Circuit Design

The DV-QNN (Fig. 1c) uses data re-uploading (Pérez-Salinas et al. 2020) with four encoding–variational layers. The 30-dimensional CNN feature vector is normalized per sample, mapped to bounded angles, and distributed across N qubits as $R_Y(\theta_i)$. Each variational layer applies $R_Z + R_Y$ rotations (2 trainable parameters/qubit/layer) followed by a nearest-neighbor CNOT chain. All qubits are measured ($2^N$-dimensional computational-basis probability) for $N \in \{3, 4, 8\}$. Two design choices proved essential for trainability and are reported as findings (Section 5.4): (i) the pre-encoding feature transformation is deterministic per sample (no dropout, per-sample rather than per-batch normalization), and (ii) variational angles are bounded to a regime where the encoding is neither near-identity nor over-rotated.

### 3.5 Training Protocol

All heads are trained end-to-end with the quantum circuit in the loop: circuit measurements feed the classification loss, whose gradients update both the classical interface and the variational gate parameters (Fig. 1d). All models: Adam (lr = 0.001, $\beta_1$ = 0.85, gradient clipping 1.0), warmup, ReduceLROnPlateau, early stopping (patience 10), L2 regularization, batch size 32, max 70 epochs, seed 42. For CV-4m and all DV sizes we report the mean ± standard deviation over three independent runs (exploiting framework non-determinism at fixed seed); CV-3m/8m and the classical model are single runs. Run-to-run variability is discussed in Section 5.7.

The quantum heads are simulated classically through PennyLane (Bergholm et al. 2022). The CV-QNN uses the Strawberry Fields Fock backend (Killoran et al. 2019b) (qumode Fock-space simulation at the stated cutoff), while the DV-QNN uses PennyLane's state-vector qubit simulator. The classical backbone and all non-quantum layers are implemented in the standard deep-learning stack and trained end-to-end with the quantum head in the loop.

### 3.6 IBM Quantum Hardware Validation

Trained DV-QNN circuits are executed on the *ibm_yonsei* backend, a 127-qubit IBM Quantum Eagle r3 processor housed in the IBM Quantum System One at Yonsei University, via Qiskit Runtime, using the Sampler primitive in Batch mode (100 test samples, 2,048 shots). CNN features and encoding angles are computed classically; only the quantum circuit runs on hardware. Hardware measurement counts are mapped to the PennyLane ordering convention before applying the trained linear readout.

### 3.7 Noise Robustness Protocol

Following Lopez et al. (2025), Gaussian noise σ is added to test images (no retraining) to probe input-noise sensitivity. This protocol is applied to a single representative model of each type, the classical CNN, the CV heads (3m, 4m), and the DV heads (3q, 4q, 8q); the analysis is qualitative, locating the noise cliff $\sigma_c$ and comparing degradation profiles across paradigms.

## 4 Results

### 4.1 Overall performance

Table 3 and Fig. 2 report test-set performance. The classical head reaches 85.0% (macro-F1 0.798). At four qumodes/qubits, matched output dimension (16) and identical readout, the CV head reaches 79.7 ± 1.8% (mean ± s.d. over three runs) while the DV head reaches 61.6 ± 1.4%, a non-overlapping 18-point gap (Fig. 2a). The CV head is also closer to classical across the ladder (78.1–81.5%), whereas the DV head sits near 60% at every size (Fig. 2b).

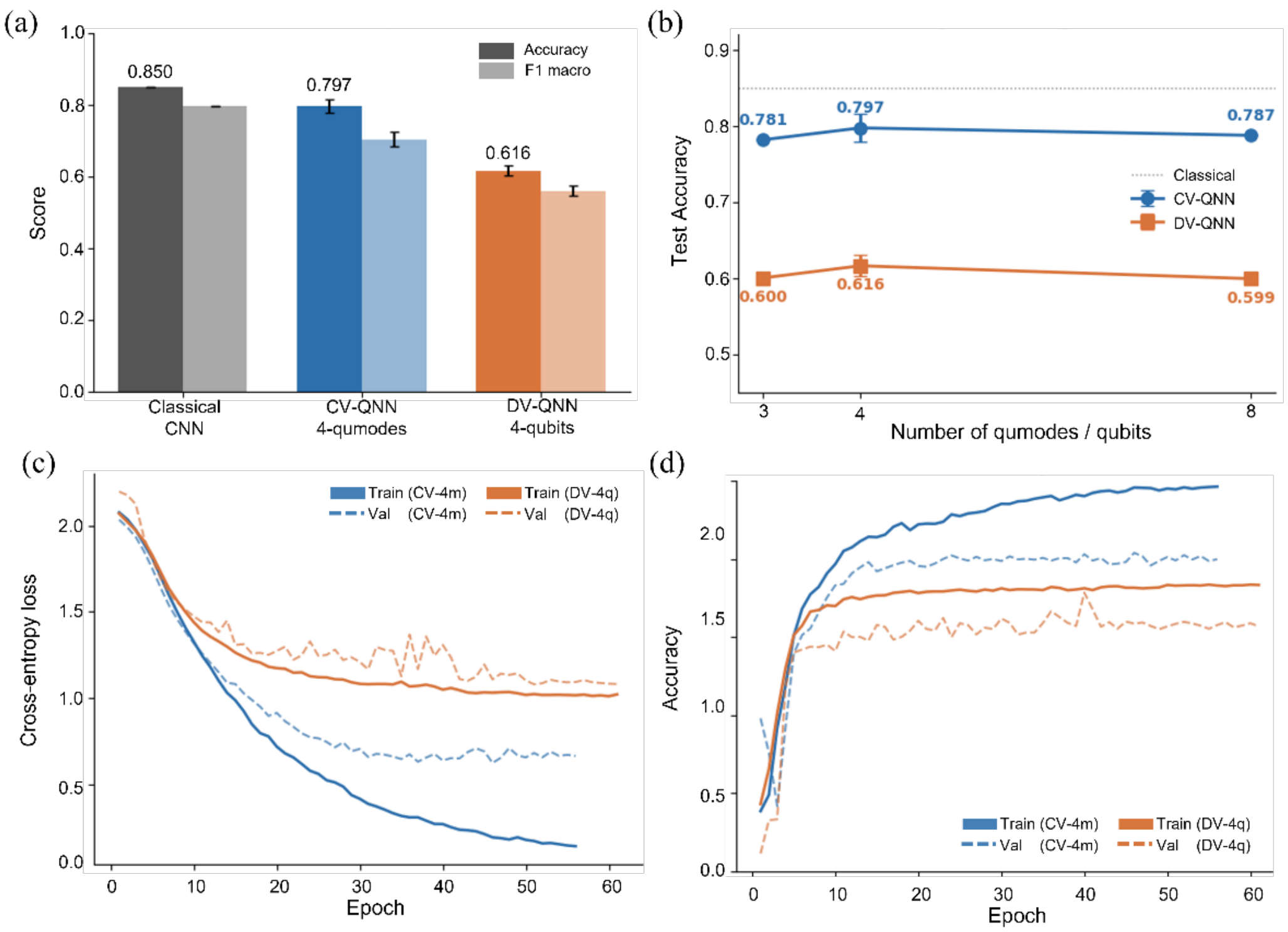

**Fig. 2 Performance and training dynamics** - (a) accuracy and macro-F1 for Classical, CV-4m, and DV-4q with quantum-parameter annotations (18-point CV–DV gap, error bars over three runs); (b) accuracy vs. size for CV (3/4/8 qumodes) and DV (3/4/8 qubits), showing CV consistently above DV and the DV head's flat ~0.60 profile. (c, d) Training dynamics for representative CV-4m and DV-4q runs: (c) cross-entropy loss and (d) accuracy versus epoch for training (solid) and validation (dashed). The DV-4q training accuracy saturates near 0.73 (loss near 1.0) while CV-4m reaches 0.99 (loss 0.14, validation 0.82), showing that the DV limitation is a representational-capacity ceiling—the model cannot fit even its training data well—rather

than an optimization failure.

A notable feature of the DV head is that its overall accuracy is itself *consistent*: 0.600 ± 0.001 (3q), 0.616 ± 0.014 (4q), and 0.599 ± 0.005 (8q), essentially flat across a four-fold change in qubit count (Fig. 2b). Adding qubits neither helps nor hurts overall accuracy, indicating that the DV head's limitation is not a matter of capacity but of representational structure. This structural reading is reinforced by the training dynamics (Fig. 2c, d): the DV head's *training* accuracy saturates near 0.73 while the CV head reaches 0.99, so the DV ceiling appears already during fitting rather than only at test time. We note that some single-run values (e.g., a four-qubit run reaching 0.75) proved to be high-variance outliers not reproduced under repeated training; all numbers reported here are run-averaged.

**Table 3.** Test-set performance (WM-811K, 2,552 samples; simulation, seed 42). CV-4m and all DV entries are mean ± s.d. over three runs; CV-3m/8m and Classical are single runs. Two per-class recalls are reported: Edge-Loc, the spatially structured class on which the paradigms diverge most, and Loc, an intrinsically hard class that no head solves (included as a control showing the CV advantage is specific to spatially structured patterns, not all hard classes). DV-QNN uses the deterministic per-sample encoding interface required for stable DV training.

| Model | Accuracy | F1 (macro) | Edge-Loc recall | Loc recall |
|---|---|---|---|---|
| Classical CNN | 0.85 | 0.798 | 0.834 | 0.567 |
| CV-QNN 3m | 0.781 | 0.674 | 0.644 | 0.35 |
| CV-QNN 4m | 0.797 ± 0.018 | 0.720 ± 0.017 | 0.661 ± 0.058 | 0.509 ± 0.055 |
| CV-QNN 8m | 0.787 | 0.711 | 0.59 | 0.469 |
| DV-QNN 3q | 0.600 ± 0.001 | 0.409 ± 0.005 | 0.042 ± 0.002 | <0.04 |
| DV-QNN 4q | 0.616 ± 0.014 | 0.560 ± 0.012 | 0.049 ± 0.020 | 0.011 |
| DV-QNN 8q | 0.599 ± 0.005 | 0.530 ± 0.011 | 0.004 ± 0.004 | <0.04 |

## 4.2 Per-class behavior: the CV advantage on spatial classes

The clearest separation between paradigms is on the spatially localized Edge-Loc class (Fig. 3). Averaged over three runs, the CV head recovers Edge-Loc with recall 0.66 ± 0.06, whereas the DV head essentially fails it at every size: 0.042 ± 0.002 (3q), 0.049 ± 0.020 (4q), and 0.004 ± 0.004 (8q). The non-overlapping statistics make this a robust finding rather than an artifact of a single run; indeed a single DV run that reached 0.77 was a high-variance outlier not reproduced under repeated training (Section 5.7). The confusion matrices make the mechanism concrete: the representative DV-4q model misclassifies 81% of Edge-Loc samples (421 of 519) as Scratch (Fig. 3c), whereas the CV-4m model assigns 66% correctly with only scattered confusion (Fig. 3b). The DV head does not encode the spatial-locality cue that distinguishes Edge-Loc from a sparse Scratch pattern.

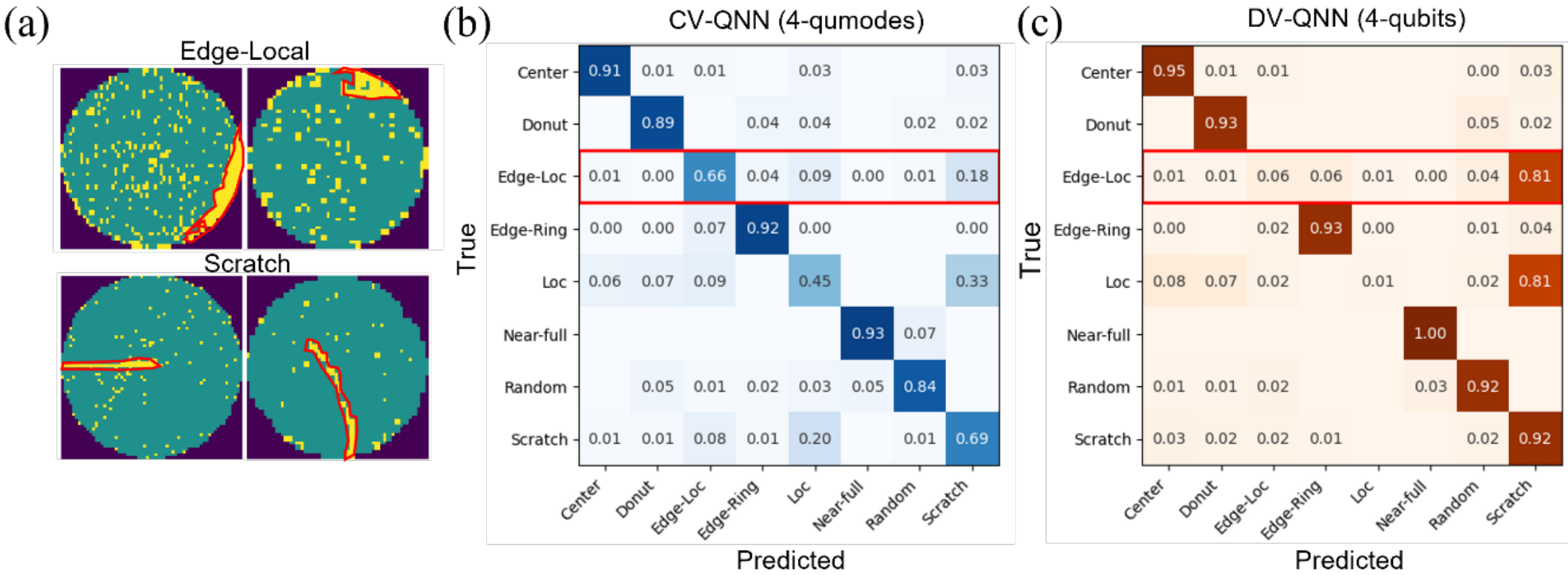

**Fig. 3 Confusion matrices, CV-4m vs. DV-4q** - (a) representative wafer maps from the Edge-Loc (top) and Scratch (bottom) classes, with defect regions outlined in red: Edge-Loc defects form a localized arc at the wafer edge whereas Scratch defects form a linear trail, a fine spatial distinction the two classes are easily confused over. (b, c) Row-normalized confusion matrices for a representative (best-validation) run of (b) CV-4m and (c) DV-4q; the accuracy shown in each title is the corresponding three-run mean from Table 3 (0.797 and 0.616), of which these runs are representative. The Edge-Loc row (highlighted) is the decisive contrast: CV concentrates on the diagonal (recall 0.66), while DV misclassifies the large majority of Edge-Loc samples as Scratch (recall ≤0.05). The same row also shows DV's near-total failure on the Loc class.

This Edge-Loc gap is the per-class signature of the overall CV advantage. Because the DV head's accuracy is flat (~0.60) while its Edge-Loc recall stays near zero across a four-fold qubit increase, the limitation is structural rather than one of capacity: adding qubits does not teach the generic DV ansatz to represent spatial locality. The Loc class serves as a control that bounds this claim: it is intrinsically hard for every head (classical recall 0.567, CV-4m 0.51 ± 0.06, DV ≤0.04 at all sizes), so the CV advantage is not a blanket improvement on all difficult classes but is specific to spatially structured patterns such as Edge-Loc. We therefore report Edge-Loc recall (where the paradigms diverge) and Loc recall (where none succeeds) side by side in Table 3: the contrast between them is what isolates spatial locality as the discriminating factor, and we claim no advantage on Loc.

A natural question is whether this confusion is intrinsic to the Edge-Loc/Scratch pair or specific to the DV encoding. The shared-backbone protocol provides a controlled test: the same backbone produces the representation that feeds all three heads, which differ only in how they subsequently encode and process it. The classical head recovers Edge-Loc with recall 0.83, confirming that the backbone features carry sufficient discriminative information; the CV head also succeeds (0.66), whereas the DV head collapses Edge-Loc almost entirely into Scratch. The failure is therefore attributable neither to the feature extractor nor to an intrinsic visual ambiguity, but to the quantum-encoding stage. Because Edge-Loc and Scratch share macroscopic morphology, the features that separate them are carried by subtle, continuous-valued amplitude differences in the backbone output. The CV displacement encoding maps these amplitudes directly into phase-space coordinates and preserves the fine-grained distinction, whereas the DV angle encoding followed by projective measurement onto the computational basis discretizes and flattens precisely these continuous cues. This flattening is later corroborated on hardware, where the DV circuit's measured outputs concentrate into a narrow region of the computational basis.

## 4.3 Input-noise robustness of the CV head

To probe sensitivity to input perturbation, we add Gaussian noise of standard deviation σ to the test images (no retraining) and measure accuracy. This analysis uses a single representative model per type, evaluated as a self-contained noise sweep; the σ = 0 baselines below are therefore those of the specific models used here and differ slightly from the run-averaged values in Table 3. We focus on the two CV cutoffs, 4-qumode (cutoff 2) and 3-qumode (cutoff 3), because the noise response is governed by the Fock cutoff rather than the qumode count; the 8-qumode model shares cutoff 2 with the 4-qumode model and so adds no new cutoff regime. Under increasing σ (Table 4, Fig. 4a), the classical CNN degrades gracefully, whereas the CV head exhibits a sharp cliff: CV-4m falls from 0.773 at σ = 0.10 to 0.453 at σ = 0.15 and 0.055 at σ = 0.20, an ~93% relative loss across Δσ = 0.10, with the steepest single drop already underway by σ = 0.10–0.15. This abrupt collapse is a robustness characteristic of the CV encoding, independent of the accuracy comparison of Section 4.1. The 3-qumode model (cutoff 3) collapses more gradually (0.223 at σ = 0.20 vs. 0.055 for 4m), an observation we revisit in Section 5.2 in light of the cutoff analysis.

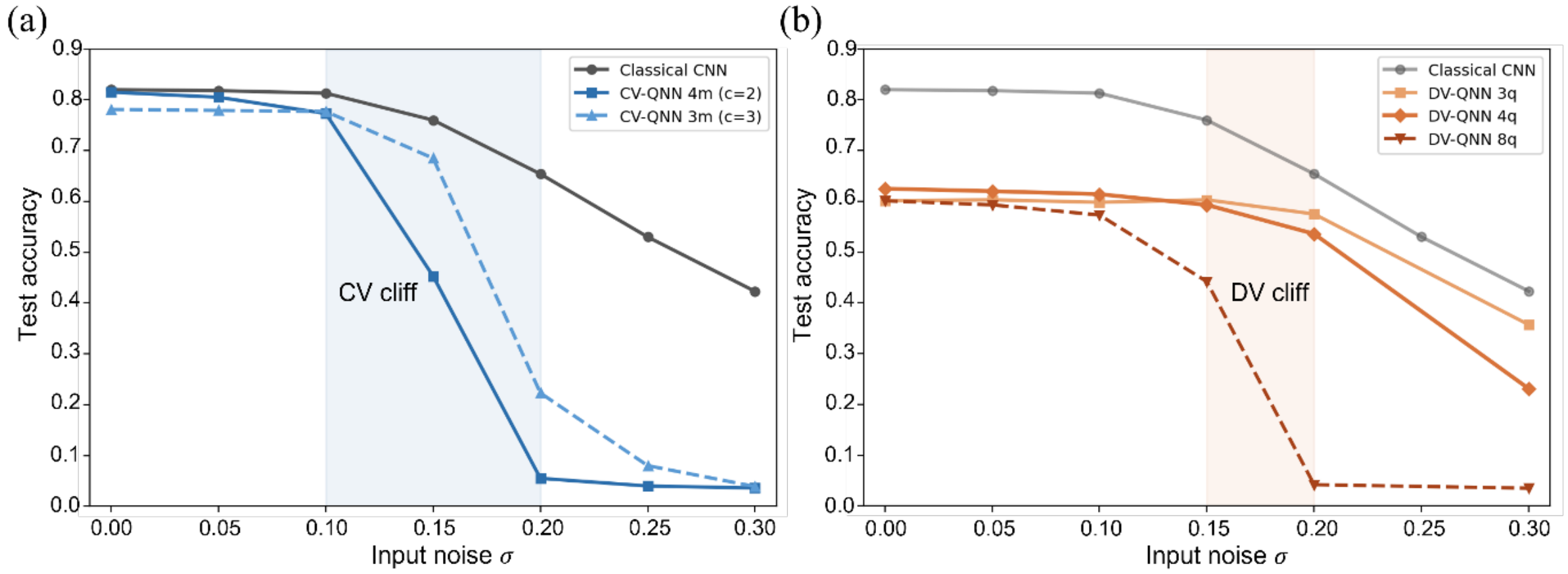

**Fig. 4 Input-noise robustness** - test accuracy under Gaussian input noise σ. (a) CV-QNN (4m, 3m) and the classical CNN: CV reaches high clean accuracy but collapses sharply over σ ≈ 0.10–0.20 (CV-4m falls from 0.773 to 0.055), while the cutoff-3 model (3m) degrades more gradually than cutoff-2 (4m). (b) DV-QNN (3q, 4q, 8q) and the classical CNN: the smaller DV heads (3q, 4q) degrade gradually from a lower baseline, which is more robust than CV-4m but at lower accuracy, whereas DV-8q shows a cliff at σ ≈ 0.15–0.20. Neither paradigm is uniformly more robust.

The DV heads show a complementary pattern (Fig. 4b). The DV-3q and DV-4q heads are in fact more robust to input noise than CV-4m: their accuracy is nearly flat through σ = 0.15 and declines only gradually thereafter (DV-4q: 0.593 at σ = 0.15, 0.536 at σ = 0.20, 0.231 at σ = 0.30), where CV-4m has already collapsed. This is a trade-off rather than a CV weakness: the DV heads maintain their accuracy under noise, but that accuracy is low (~0.60) to begin with, below the classical baseline at every σ, so their robustness is robustness of an already weaker classifier. The DV-8q head is the exception, showing a CV-like cliff (0.442 at σ = 0.15 to 0.042 at σ = 0.20); the deepest circuit is the most noise-sensitive, mirroring its hardware behavior. Taken together, neither paradigm is uniformly more robust: CV reaches higher accuracy but degrades abruptly, the smaller DV heads degrade gently from a lower level, and the largest

DV head inherits a cliff of its own.

**Table 4.** Accuracy under Gaussian input-noise injection (selected σ; single representative model per type; σ = 0 baselines are model-specific and differ from the Table 3 means). CV and DV sweeps use independent representative models.

| σ | Classical | CV-QNN 3m | CV-QNN 4m | DV-QNN 3q | DV-QNN 4q | DV-QNN 8q |
|---|---|---|---|---|---|---|
| 0 | 0.82 | 0.781 | 0.815 | 0.601 | 0.625 | 0.601 |
| 0.1 | 0.813 | 0.777 | 0.773 | 0.598 | 0.614 | 0.573 |
| 0.15 | 0.76 | 0.685 | 0.453 | 0.603 | 0.593 | 0.442 |
| 0.2 | 0.654 | 0.223 | 0.055 | 0.575 | 0.536 | 0.042 |
| 0.3 | 0.423 | 0.039 | 0.036 | 0.357 | 0.231 | 0.035 |

### 4.4 DV-QNN validation on IBM quantum hardware across circuit depths

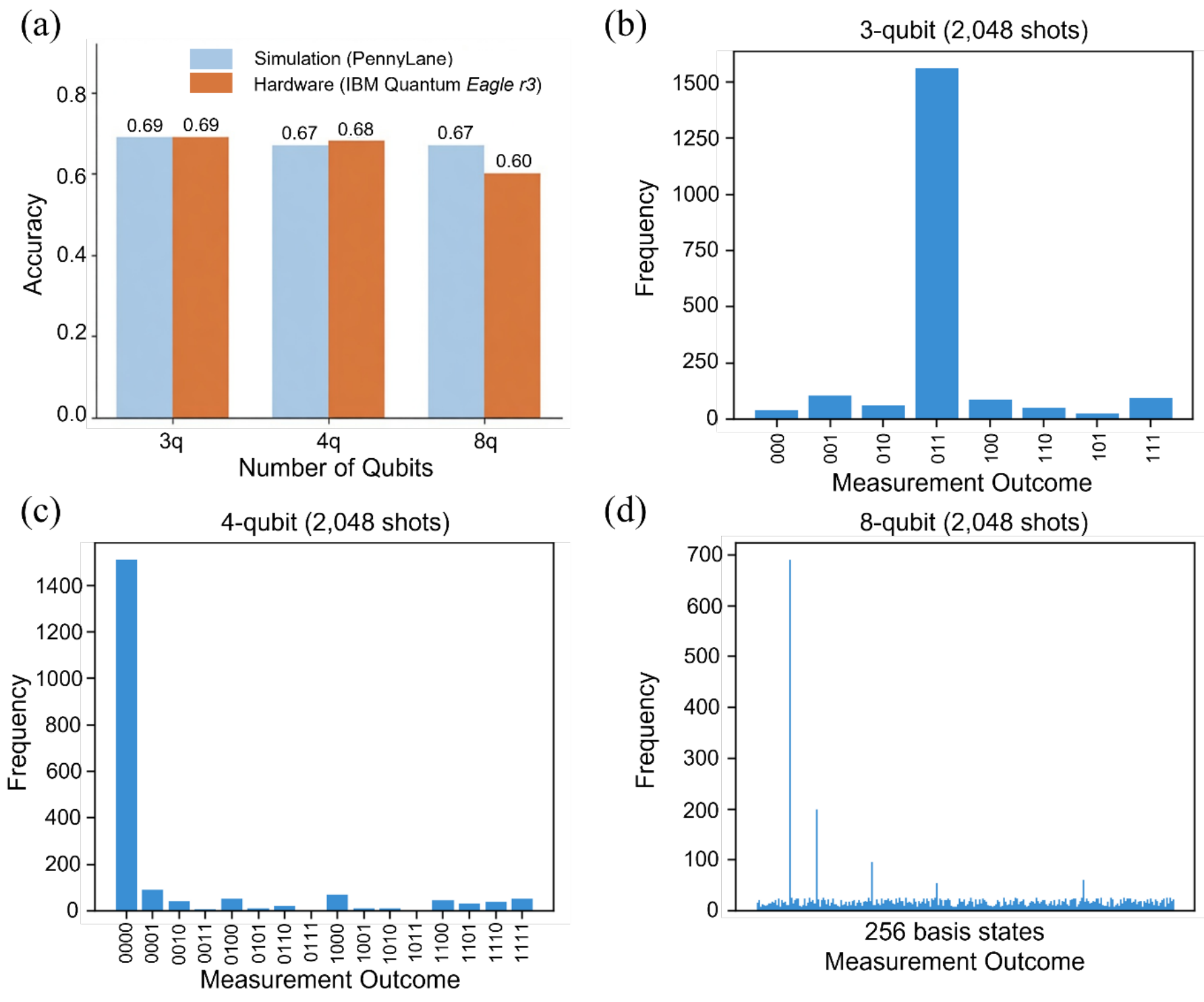

**Fig. 5 IBM quantum hardware validation** - (a) simulation vs. hardware accuracy for DV-QNN at three depths (*ibm_yonsei*, Eagle r3; 100-sample subset, 2,048 shots): accuracy is preserved through 12 two-qubit gates (3q, 4q; $\Delta \approx +0.01$) and degrades by 7 points at 28 gates (8q, $\Delta \approx -0.07$). (b–d) Measured output distributions (2,048 shots) for the 3-, 4-, and 8-qubit circuits: probability concentrates on a single basis state for 3q ('011') and 4q ('0000'), and remains markedly non-uniform, though more dispersed, for 8q, consistent with the DV head mapping inputs into a narrow region of the computational basis.

Table 5 and Fig. 5a compare ideal simulation and *ibm_yonsei* hardware on an identical 100-sample subset, using the representative (best-validation) model at each size. At shallow depth the hardware accuracy matches simulation to within sampling resolution (3q: 0.69 vs 0.69; 4q: 0.68 vs 0.67), with the +0.01 differences consistent with shot noise. At the deepest circuit (8 qubits, 28 transpiled two-qubit gates) the hardware accuracy falls by 7 points (0.60 vs 0.67).

This depth-resolved pattern—accuracy preserved through 12 two-qubit gates and degrading at 28—gives a concrete, monotone picture of NISQ noise acting on the trained arg-max decision. The measured output distributions (Fig. 5b–d) make the encoding behavior visible directly: probability concentrates on a single basis state for the 3- and 4-qubit circuits ('011' and '0000') and remains markedly non-uniform, though more dispersed, for the 8-qubit circuit. On hardware, the deeper 8-qubit circuit's larger gate count further erodes the (already weak) class structure, consistent with the more dispersed measurement distribution seen for 8 qubits (Fig. 5d).

**Table 5.** Simulation vs. IBM hardware (*ibm_yonsei*; identical 100-sample subset, 2,048 shots; representative models).

| Model | Accuracy (Simulation) | Accuracy (IBM Hardware) | Difference (Δ) | Two-qubit gates | Transpiled circuit depth |
|---|---|---|---|---|---|
| DV-QNN 3q | 0.69 | 0.69 | +0.01 | 8 | 38 |
| DV-QNN 4q | 0.67 | 0.68 | +0.01 | 12 | 43 |
| DV-QNN 8q | 0.67 | 0.6 | −0.07 | 28 | 52 |

## 5 Discussion

### 5.1 The CV advantage: structured design and continuous encoding

The CV head outperforms the DV head decisively and consistently, by 18 points in accuracy and, most tellingly, on the spatially localized Edge-Loc class (0.66 vs 0.05). We argue this advantage stems from two properties of the CV head that operate together, rather than from the size of the Hilbert space (a distinction we make precise in Section 5.2): a structured, neural-network-analogue layer design, and a continuous, real-valued phase-space encoding. Both are intrinsic to the CV paradigm.

*Structured design.* A Killoran CV layer is an explicit quantum analogue of a neural-network layer: an interferometer–squeezer–interferometer block implements an affine map of the encoded quadratures, a displacement supplies the bias, and a non-Gaussian gate supplies the nonlinearity. This "linear-map-then-nonlinearity" structure is precisely the inductive bias gradient-based training is built around. The DV head, by contrast, is a generic hardware-efficient ansatz with no comparable correspondence to a neural-network layer, and its accuracy is essentially flat (~0.60) from three to eight qubits: quadrupling the Hilbert space neither helps nor hurts. The training curves make this structural reading explicit, as we detail below.

*Continuous encoding.* The per-class evidence points to a second, complementary factor. The decisive failure is on Edge-Loc, which the representative DV head misclassifies as Scratch 81% of the time (Fig. 3c), two patterns that differ by a fine spatial distinction (a localized edge arc versus a sparse line). A continuous, real-valued phase-space representation can preserve such graded spatial cues, whereas the discrete angle encoding of the DV head appears to flatten them: the DV head learns the easier ring- and full-pattern classes well but cannot represent the spatial-locality feature that separates Edge-Loc from Scratch, at any circuit size. We therefore read the CV/DV contrast as the difference between a structured, continuously encoded circuit and an unstructured, discretely encoded one, and we locate the practical value of the CV paradigm in this combination.

We note that our design does not separate these two factors, structure and continuous encoding co-vary between the CV and DV heads, so we do not claim to isolate their individual contributions; disentangling them (for example, by a continuously encoded but unstructured ansatz) is left for future work. We further caution that the DV head is high-variance run to run: an isolated four-qubit run reached 0.75 with Edge-Loc 0.77, but this was not reproduced, and the run-averaged behavior is the flat, spatial-blind profile above.

The training curves (Fig. 2c, d) make the nature of the DV limitation explicit and rule out an "under-training" explanation. The DV-4q head does not stall in optimization—its training loss descends steadily—but its training accuracy saturates near 0.73 and its loss plateaus near 1.0, i.e., the model fits its own training data only to a low ceiling. The CV-4m head, by contrast, drives training accuracy to 0.99 (loss to 0.14) and reaches 0.82 validation accuracy. The gap is therefore not one of generalization alone: the DV circuit cannot represent the training set well in the first place, which is the signature of a representational-capacity ceiling rather than a barren-plateau or optimization failure (whose mitigation, e.g., by tailored initialization, is itself an active area; Grant et al. 2019). This directly supports the structural reading: the limitation is in what the unstructured ansatz can express, not in whether it can be optimized.

## 5.2 Depth-resolved hardware behavior and the CV noise cliff

The CV-QNN literature often attributes the expressive power of CV models to an infinite-dimensional Fock space and to non-Gaussian resources from the Kerr gate (Killoran et al. 2019a; Choe 2022). Neither applies at the cutoff used here. With Fock cutoff d = 2, each qumode is truncated to span$\{|0\rangle, |1\rangle\}$, whose dimension equals that of a single qubit; the "infinite-dimensional" description does not apply to the simulated model. Moreover, $K(\kappa)|n\rangle = e^{ikn^2}|n\rangle$ acts on this space as diag$(1, e^{ik})$, since $n^2 = n$ for $n \in \{0, 1\}$. This is a diagonal phase rotation, a Gaussian-equivalent single-qumode operation, so at d = 2 the Kerr gate supplies no non-Gaussian resource, and the model's behavior cannot be ascribed to breaking the Gaussian barrier in the sense of the efficient classical description of Gaussian/stabilizer operations (Gottesman 1998; Bartlett et al. 2002). The CV head's advantage is therefore better understood as a property of (i) its continuous, real-valued phase-space encoding and (ii) its structured affine-plus-nonlinearity layer design, not of Hilbert-space dimensionality or non-Gaussian universality. Genuine non-Gaussian effects require $d \geq 3$ (so that $n^2 \neq n$); our limited d = 3 (three-qumode) data are more, not less, robust to injected noise (Section 4.3); since noise robustness and non-Gaussianity are distinct axes, this single observation neither establishes nor refutes a non-Gaussian contribution, and we present it only as not providing positive evidence for one. Relatedly, the universality of the CV affine map requires a full interferometer mesh; the reduced beamsplitter cascade used in practice realizes only a subset of orthogonal transformations for $m > 4$, so any claim of an "arbitrary W" holds only subject to this restriction.

## 5.3 Interpreting the noise results: hardware tolerance and the CV cliff

The 3/4/8 ladder gives a depth-resolved picture of hardware behavior: arg-max classification accuracy is preserved through 12 transpiled two-qubit gates (3q and 4q, $\Delta \approx +0.01$, within shot-noise resolution) and degrades by 7 points only at the deepest circuit (8q, 28 two-qubit gates). Because simulation and hardware use the identical representative model and 100-sample subset, this $\Delta$ reflects hardware noise alone. The result is useful for hybrid classifiers specifically: the trained readout depends only on the arg-max outcome, which tolerates the noise of shallow circuits but begins to fail as the two-qubit gate count grows. Separately, the CV input-noise cliff must be interpreted in light of the cutoff caveat of Section 5.2: at $d = 2$ there is no quadratic Kerr term ($n^2 = n$) to amplify perturbations, so the steep degradation is better attributed to the squeezing gate's exponential quadrature scaling, $S(r)$ compresses one quadrature ($\hat{x} \rightarrow e^{-r}\hat{x}$) but anti-squeezes the orthogonal one ($\hat{p} \rightarrow e^{r}\hat{p}$), so input perturbations along the amplified quadrature are exponentially magnified, and to truncation dynamics, rather than to Kerr nonlinearity.

### 5.4 The classical–quantum encoding interface as a design variable

A concrete, transferable finding emerged from making the DV head trainable. A standard classical regularizer, dropout placed immediately before the quantum encoding, silently destroyed DV training: by perturbing the feature vector on every forward pass it rendered the encoding angles non-stationary, and the low-parameter quantum circuit could not track a moving target, collapsing to majority-class prediction. A near-identity encoding (very small rotation angles) likewise failed, by making the circuit nearly input-independent. Removing pre-encoding dropout, normalizing features per sample rather than per batch (batch normalization injects a second, batch-dependent form of encoding noise), and bounding the encoding angles away from both the near-identity and over-rotated regimes were jointly necessary and sufficient to recover DV training. This is an instance of the general point in Section 5.1: the unstructured DV head is far more sensitive to the design of the classical–quantum interface than the structured CV head, and hybrid pipelines should treat the encoding boundary as deterministic.

### 5.5 Process-level implications of the Edge-Loc recall gap

Beyond the classification metrics, the Edge-Loc recall gap carries a concrete manufacturing consequence. When a DV-based classifier labels an Edge-Loc wafer as Scratch, an automated disposition system would flag the CMP module or transfer-robot alignment for inspection, neither of which is the actual source. The true root cause, load-lock valve contamination or thermal-annealing drift (Table 1), would go unaddressed and could propagate across subsequent lots. The CV head's recall of 0.66, although still below the classical baseline, correctly routes two-thirds of Edge-Loc wafers to the thermal-process domain.

This illustrates a broader point for wafer-map classification: per-class recall on morphologically ambiguous classes can matter more for yield management than aggregate accuracy. A classifier with high aggregate accuracy that systematically misroutes one ambiguous defect type may inflict greater cumulative yield loss than one with slightly lower accuracy whose errors are more evenly distributed. The CV paradigm's ability to preserve the continuous spatial cues that separate such classes, even at the modest Fock cutoff $d = 2$, where the Kerr gate is Gaussian-equivalent, suggests that continuous-variable encoding may be well suited to industrial classification tasks in which fine-grained spatial information is decision-critical.

### 5.6 Parameter counts and comparison with prior work

We avoid framing the comparison as one of parameter efficiency. Although the CV and DV circuits use few trainable parameters (190 and 32 at four qumodes/qubits), each requires a classical interface layer that dominates the head's parameter budget; the quantum heads are therefore not more parameter-efficient than the classical head at the head level. Our controlled design holds total capacity fixed and varies only the head mechanism, so parameter count is not the comparison axis. Relative to Yang and Sun (2022), whose SP&A-Net backbone reaches 98.10%, our deliberately simple backbone answers a different question, the intrinsic comparison of CV and DV heads, rather than competing on absolute accuracy.

### 5.7 Limitations and Future Work

Several limitations bound the scope of these findings. First, on replication: accuracies are averaged over three runs for CV-4m and all DV sizes (seed 42, exploiting framework non-determinism), but CV-3m/8m and the classical model are single runs and all use a single seed. The same single-run caution we apply to outlier DV runs therefore applies to the CV-3m/8m endpoints, so the "consistent across the ladder" reading rests most firmly on the three-run CV-4m point. The DV head is high-variance: an isolated four-qubit run reached 0.75 (Edge-Loc 0.77) but was not reproduced, with the three-run mean at 0.62 (Edge-Loc 0.05), so single-run DV numbers should be treated with caution. Broader multi-seed replication would strengthen the size-scaling claims. Second, on scope: the hardware evaluation covers one backend, one execution, 100 samples, and no error mitigation (Temme et al. 2017), and the CV results are from classical simulation only, so photonic-hardware behavior may differ. Third, on the backbone: a deliberately simple feature extractor was used to isolate the quantum head, so absolute accuracies fall below classical state of the art and are not directly comparable to heavy-backbone results (Yang and Sun 2022). Two further caveats are developed where they arise rather than repeated here: the finite Fock cutoff $d = 2$, at which the CV model is not genuinely non-Gaussian (Section 5.2), and the Loc class, which no head solves and on which we therefore claim no quantum advantage (Section 4.2).

These bounds point to concrete next steps: multi-seed and multi-backend replication with error mitigation; extension to higher Fock cutoffs ($d \geq 3$), where genuine non-Gaussian resources become available and the two pillars of the CV advantage (structured design and continuous encoding) could be disentangled by an ablation with a continuously encoded but unstructured ansatz; and evaluation with stronger backbones to test whether the CV head's per-class advantage on spatially structured defects persists closer to the state of the art.

## 6 Conclusion

Through a controlled comparison on WM-811K, we find that the CV quantum head consistently outperforms the DV head, by 18 points in accuracy (79.7% vs 61.6% at four qumodes/qubits, over three runs) and decisively on the spatially localized Edge-Loc class (recall 0.66 vs ≤0.05). The DV head's accuracy is flat across a four-fold change in qubit count and its Edge-Loc recall stays near zero, indicating a structural rather than capacity limitation. We attribute the CV advantage to two intrinsic properties of the CV head, its principled neural-network-analogue construction and its continuous phase-space encoding, rather than to DV's generic ansatz. We also clarify the basis of this advantage: at the cutoff used here it stems from structured design and continuous encoding, not from Hilbert-space dimensionality or non-Gaussian resources as is sometimes assumed. On hardware, DV-QNN classification accuracy is preserved through 12 transpiled two-qubit gates and degrades at 28, giving a clean depth-resolved view of NISQ noise; and DV trainability hinges on a deterministic classical–quantum interface. These results offer design guidance for hybrid quantum heads in industrial imaging: prefer structured, neural-network-analogue CV layers, and treat the encoding boundary as deterministic.

## Declarations

**Funding:** -

**Data availability:** The WM-811K wafer-map dataset is publicly available (Wu et al. 2015). The processed data splits and trained model checkpoints supporting the findings of this study will be made available in a public repository upon publication.

**Code availability:** The code used to train the models and generate the figures will be released in a public repository upon publication.

**Author contributions:** Y.K. and K.K. conceived the study and designed the controlled comparison protocol. Y.K. designed the quantum circuits, implemented the models, performed the simulations(Pennylane, StrawberryFields) and data analysis, and wrote the manuscript. Y.K. and J.I. executed and validated the DV-QNN circuits on IBM quantum hardware. M.B., K.K. supervised the overall research. All authors reviewed and approved the final manuscript.

**Acknowledgments:** We acknowledge the use of IBM Quantum services for this work, accessed through the IBM Quantum System One at Yonsei University. The views expressed are those of the authors and do not reflect the official policy or position of IBM or the IBM Quantum team.